\documentclass[conference]{IEEEtran}
\IEEEoverridecommandlockouts
\usepackage{cite}
\usepackage{stfloats}
\usepackage{amsmath,amssymb,amsfonts}
\usepackage{algorithmic}
\usepackage{graphicx}
\usepackage{textcomp}
\usepackage{xcolor}
\def\BibTeX{{\rm B\kern-.05em{\sc i\kern-.025em b}\kern-.08em
		T\kern-.1667em\lower.7ex\hbox{E}\kern-.125emX}}
\IEEEoverridecommandlockouts
\IEEEpubid{\makebox[\columnwidth]{978-1-7281-1025-7/19/\$31.00 ©2019 IEEE\hfill} \hspace{\columnsep}\makebox[\columnwidth]{ }}

\begin{document}
	
	\title{A Method Based on Hierarchical Spatiotemporal Features for Trojan Traffic Detection\\
	}
	
	\author{
		\IEEEauthorblockN{Jiang Xie\IEEEauthorrefmark{1}\IEEEauthorrefmark{3},Shuhao Li\IEEEauthorrefmark{1}\IEEEauthorrefmark{2},Yongzheng Zhang\IEEEauthorrefmark{1}\IEEEauthorrefmark{2}\IEEEauthorrefmark{3},Xiaochun Yun\IEEEauthorrefmark{4},Jia Li\IEEEauthorrefmark{1}\IEEEauthorrefmark{3}}
		
		\IEEEauthorblockA{\IEEEauthorrefmark{1}Institute of Information Engineering, Chinese Academy of Sciences, Beijing, China}
		\IEEEauthorblockA{\IEEEauthorrefmark{2}	Key Laboratory of Network Assessment Technology, IIE, CAS, Beijing, China}
		\IEEEauthorblockA{\IEEEauthorrefmark{3}School of Cyber Security, University of Chinese Academy of Sciences, Beijing, China}
		\IEEEauthorblockA{\IEEEauthorrefmark{4}National Computer Network Emergency Response Technical Team/Coordination Center of China, Beijing, China}
		\IEEEauthorblockA{Email: \{xiejiang,lishuhao,zhangyongzheng\}@iie.ac.cn; \{yunxiaochun,lijia\}@cert.org.cn}
		
	}
	\maketitle
	
	\begin{abstract}
		Trojans are one of the most threatening network attacks currently. HTTP-based Trojan, in particular, accounts for a considerable proportion of them. Moreover, as the network environment becomes more complex, HTTP-based Trojan is more concealed than others. At present, many intrusion detection systems (IDSs) are increasingly difficult to effectively detect such Trojan traffic due to the inherent shortcomings of the methods used and the backwardness of training data. Classical anomaly detection and traditional machine learning-based (TML-based) anomaly detection are highly dependent on expert knowledge to extract features artificially, which is difficult to implement in HTTP-based Trojan traffic detection. Deep learning-based (DL-based) anomaly detection has been locally applied to IDSs, but it cannot be transplanted to HTTP-based Trojan traffic detection directly. To solve this problem, in this paper, we propose a neural network detection model (HSTF-Model) based on hierarchical spatiotemporal features of traffic. Meanwhile, we combine deep learning algorithms with expert knowledge through feature encoders and statistical characteristics to improve the self-learning ability of the model. Experiments indicate that \({F}_{1}\) of HSTF-Model can reach 99.4\% in real traffic. In addition, we present a dataset BTHT consisting of HTTP-based benign and Trojan traffic to facilitate related research in the field.
	\end{abstract}
	
	
	\section{Introduction}
	The rapid development of the Internet has brought great convenience. However, more comprehensive service means that traffic becomes more complex, which poses greater challenges to network security. Trojans are one of the main malicious attacks in online activities. They are a class of malicious programs that attack hosts/websites to steal personal information and even remotely control devices. In 2017, a total of 19,017,282 hosts with IP addresses were implanted with Trojans. And monthly, almost 2.81 million host IP addresses in the global Internet were infected with "Flying" worm Trojan in average\cite{Xbookhulianwangchina}. These Trojans spread and enforce attacks through network traffic, where HTTP traffic is one of the main carriers. Therefore, it is necessary to find an effective way to detect HTTP-based Trojan traffic.
	
	At present, it is a highly concerned issue about ensuring the security of network devices and information. People usually build intrusion detection systems (IDSs) to resist various network attacks. The anomaly detection, which is a primary research direction in the field of intrusion detection, can detect new and unknown attacks (0-day) by analyzing benign and malicious behavior characteristics from traffic\cite{liao2013intrusion}. However, most of the current anomaly detection methods, including machine learning-based methods, are highly dependent on feature engineering to achieve high detection performance. And designing a satisfactory feature engineering that can accurately extract traffic characteristics is a hard work\cite{zhang2013effective}. Therefore, many anomaly detection methods are difficult to apply in practice\cite{hubballi2014false}.
	
	In recent years, deep learning has attracted great attention in various fields. And many researchers have proposed anomaly detection methods based on deep learning to automatically extract features\cite{yin2017deep, shone2018deep}. After inputting low-order data features, neural networks can correct model parameters to find complex structures from data, and then abstract them into high-order data to realize automatic learning. Deep learning reduces the dependence of model on feature engineering, which allows researchers to design excellent anomaly detection models through simple data preprocessing.
	
	In this paper, to effectively detect HTTP-based Trojan traffic and reduce the dependence on feature engineering, we present an anomaly detection method based on deep learning algorithms. Then, a statistical feature set based on experience is added to improve the self-learning ability of the model.
	
	In general, the contributions of this paper are as follows.
	
	\begin{itemize}
		\item We combine raw traffic with statistical characteristics to enrich feature representation information. Then, feature encoders are built to layer the traffic into packet-level and flow-level for preprocessing. Experiments indicate that this preprocessing method can effectively improve the performance of the model.
		\item We propose HSTF-Model (Model based on Hierarchical SpatioTemporal traffic Features) consisting of CNN and LSTM. HSTF-Model extracts hierarchical spatiotemporal sequence features from raw data and statistical characteristics. We can adjust the weight of the two to determine which one is more reliable.
		\item We generate a dataset BTHT-R (Benign and Trojan Traffic based on Http-Raw) consisting of raw HTTP traffic. BTHT-R includes benign traffic (4,044,751 flows) and HTTP-based Trojan traffic (37,847 flows). In addition, a corresponding statistical feature dataset BTHT-S (Statistical) is also built. BTHT-R and BTHT-S form dataset BTHT\footnote{The dataset can be found at https://drive.google.com/open?id=1d\_SVIOzz\\
			gw2kYPlC5dKjgOl51YXTDZUi. Researchers who are going to use the dataset should indicate the original source of data by citing this paper.}. Because of privacy, we use irreversible hash technology for data masking in BTHT.
	\end{itemize}
	
	The remainder of this paper is organized as follows. Related works is described in Section \uppercase\expandafter{\romannumeral2}, and the introduction of dataset BTHT in Section \uppercase\expandafter{\romannumeral3}. Section \uppercase\expandafter{\romannumeral4} introduces the methodology of HSTF-Model. We conduct experiments in Section \uppercase\expandafter{\romannumeral5}. Subsequently, we discuss the model and experiments in Section \uppercase\expandafter{\romannumeral6}. Section \uppercase\expandafter{\romannumeral7} draws our conclusion and gives future challenges.
	
	\section{Related Work}	
	The research in this paper belongs to malicious traffic detection, which is one of the cores of intrusion detection. In general, there are two types of approaches for IDSs based on traffic detection, signature detection and anomaly detection\cite{depren2005intelligent}. The former, also called misuse detection, mainly analyzes the characteristics and behavior patterns of known attacks. The greater the similarity of network behavior in actual detection, the more likely that it is to be judged malicious. Signature detection is effective in detecting known attacks with low errors. But it cannot detect new and unknown attacks (0-day). The latter is also known as behavior detection. Anomaly detection can detect unknown attacks. This makes anomaly detection a major research direction in intrusion detection\cite{bhuyan2013network}. Current mainstream anomaly detection methods include classical anomaly detection, TML-based anomaly detection, DL-based anomaly detection and so on.
	
	\subsection{Classical Anomaly Detection}
	The classical anomaly detection profiles benign traffic patterns\cite{depren2005intelligent}. It is based on the hypothesis that an attacker behavior differs to that of a benign user. Benign operations of the members are profiled and a certain amount of deviation from the benign behavior is flagged as an anomaly\cite{can2015survey}. 
	
	Classical anomaly detection can be useful for new attack patterns, but it is not as effective as signature detection when detecting known attacks\cite{kemmerer2002intrusion}. In addition, its performance is highly dependent on feature engineering. If the feature analysis for benign traffic is incomplete, this approach will have a high false-positive rate\cite{depren2005intelligent}. Therefore, more research is attempting to combine classical anomaly detection and signature detection currently, that is, analyzing both benign and malicious traffic to improve the performance of model. Because this hybrid intrusion detection can detect unknown attacks, it is also commonly considered as anomaly detection\cite{bhuyan2013network}.
	
	\subsection{TML-based Anomaly Detection}
	There are many traditional machine learning-based (TML-based) anomaly detection methods applied to IDSs\cite{buczak2015survey}, for instance, Markov-based anomaly detection\cite{chen2016anomaly}. These TML-based intrusion detection methods simultaneously learn the characteristics of benign and malicious traffic.
	
	Lin \emph{et al.}\cite{lin2015cann} presented CANN, a feature representation approach that combines cluster centers and nearest neighbors for intrusion detection. The experimental results based on the dataset KDD-Cup'99\cite{rosset2000kdd} can reach 99.9\% in accuracy. Aljawarneh \emph{et al.}\cite{aljawarneh2018anomaly} proposed a hybrid model based on the optimal characteristics of network traffic. The hybrid algorithm is composed of J48, Meta Pagging, and other classical machine learning algorithms. The accuracy of the model can be 99.81\% (binary-class) and 98.56\% (multiple-class) in dataset NSL-KDD\cite{tavallaee2009detailed}, respectively. Chen \emph{et al.}\cite{chen2018machine} proposed S-IDGC based on machine learning to classify imbalanced traffic data for mobile malware detection. Gezer \emph{et al.}\cite{gezer2019flow} utilize machine learning techniques to detect TrickBot malware infections.
	
	In general, most TML-based anomaly detection methods also need to design feature engineering. Some traffic features are designed first. Then, a model is built based on those features using supervised or unsupervised learning algorithms. But designing a feature set that can accurately characterize network traffic is still an ongoing research issue\cite{zhang2013effective}.
	
	\subsection{DL-based Anomaly Detection}
	At present, deep learning (DL) have received great attention in various fields, including cyber security\cite{Yang2018Machine}. Multiple processing layers of the neural network can learn the abstract representation of data. Neural network extracts low-order features directly from the raw data, and then, combines and transforms them into high-order features for automatic learning and analysis. Therefore, DL-based anomaly detection reduces the need for researchers to spend more time designing complex feature engineering\cite{dong2016comparison}.
	
	Shone \emph{et al.}\cite{shone2018deep} presented a novel DL technique using stacked NDAEs for intrusion detection, which is a nonsymmetric deep autoencoder for unsupervised feature learning. Accuracy can reach 97.85\% in KDD-Cup'99 and 89.22\% in NSL-KDD. CNN and RNN are two widely used neural network models. CNN has attracted much attention for its spatial feature extraction ability. In cyber security, Vinayakumar \emph{et al.}\cite{vinayakumar2017applying} used CNNs for network intrusion detection. Li \emph{et al.}\cite{li2017intrusion} used CNN for representation learning for intrusion detection. Yin \emph{et al.}\cite{yin2017deep} proposed RNN-IDS for intrusion detection, and compared it with classical machine learning algorithms. LSTM, another well-known model as one of a variant of RNN, has natural advantages in processing data with sequences and dependencies. Kim \emph{et al.}\cite{kim} applied LSTM for intrusion detection, which can achieve accuracy 96.93\%. 
	
	On the basis of existing research, there are the following main problems that can be improved: 1) The classic datasets for benchmarking, such as KDD-Cup'99, are outdated. The actual network environment is more complex nowadays. 2) There is no specific research for the detection of the HTTP-based Trojan traffic attack scenario. 3) Classical and TML-based anomaly detection methods are mostly dependent on well-designed feature engineering. DL-based anomaly detection methods can extract features automatically, but most of those methods currently learn directly from the original data, without making use of artificial accumulated feature experience. Our method belongs to DL-based anomaly detection. Meanwhile, we add statistical characteristics based on experience and data analysis to get a more efficient neural network model.
	
	\section{Feature Set Construction}
	In this paper, we generate dataset BTHT (Benign and Trojan traffic based on Http) including BTHT-R (Raw) and BTHT-S (Statistical). BTHT-R consists of benign and malicious traffic from the laboratory gateway and traffic interfaces of CNCERT/CC\footnote{CNCERT/CC is the abbreviation of “National Computer Network emer- gency Response technical Team/Coordination Center of China”. The web site of CNCERT/CC is http://www.cert.org.cn/.}. For each flow in BTHT-R, statistical characteristics are made to construct dataset BTHT-S.
	
	All the data used in experiments is from BTHT. We are also set up multiple proportions of benign:malicious data to evaluate the performance of HSTF-Model comprehensively.
	
	\subsection{Data Acquisition}
	Dataset BTHT-R is generated by capturing and filtering traffic from the real network. There are about 4 million flows in BTHT-R, 99\% of which come from benign behavior and 1\% come from malicious behavior (various Trojans). 
	
	For most benign traffic, after being authorized, we collected about 300GB traffic from the laboratory gateway under the premise of protecting privacy. After cleaning out irrelevant/redundant information in traffic, we obtained 4,044,751 million benign network flows eventually, including news browsing, social chatting, web browsing and so on.
	
	HTTP-based Trojan traffic is provided by CNCERT/CC. There are two sources of HTTP-based Trojan traffic. One is to analyze the real-time traffic in the network through the existing monitoring system of CNCERT/CC. The other is to deploy honeypots in the network for malicious sample capture and breeding to generate HTTP-based Trojan traffic. After analysis and processing, we obtained 37,847 valid malicious flows based on HTTP by manual labeling. These malicious traffic types include malicious promotion, malicious download, Trojans implant and so on. Because of data security protection, the sensitive fields (host, etc.) of traffic we obtained were subjected to irreversible hash processing. To maintain consistency, benign traffic is handled in the same way.
	
	\vspace{-1em}
	\begin{figure}[htbp]
		\centerline{\includegraphics[scale=0.4]{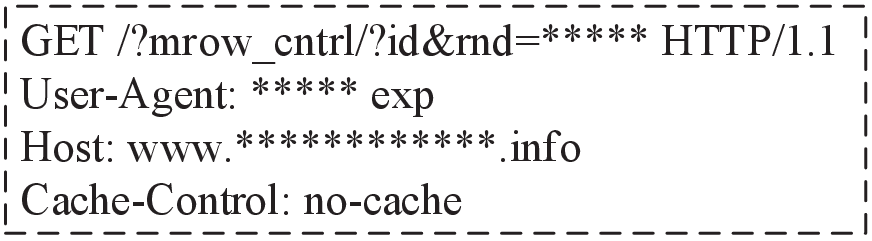}}
		\vspace{-0.5em}
		\caption{Online package generated by a Trojan intrusion.}
		\label{fig1}
	\end{figure}
	\vspace{-0.5em}
	\newcommand{\tabincell}[2]{\begin{tabular}{@{}#1@{}}#2\end{tabular}} 
	
	An online package generated by a Trojan intrusion behavior is shown visually in Fig.~\ref{fig1}. The relevant sensitive data is replaced with '*' to fully protect the privacy and data security. In addition, data size in BTHT-R is shown in Tab~\ref{tab2}.
	
	\vspace{-1em}
	\begin{table}[htbp]
		\caption{Statistics on packet size (in bytes) and flow size (in packets) in BTHT-R}
		\vspace{-1em}
		\begin{center}
			\begin{tabular}{|c|c|c|}
				\hline
				\textbf{Statistics}&\textbf{Packet} &\textbf{Flow}\\
				\hline 
				\hline
				Count & 14,892,047 & 4,082,598 \\
				\hline
				Size & 12,597,103,235 & 14,892,047 \\
				\hline
				Mean & 845.894 & 3.648 \\
				\hline
				Min & 12 & 1\\
				\hline
				Max & 46,729 & 16,819 \\
				\hline
			\end{tabular}
			\label{tab2}
		\end{center}
		\vspace{-2em}
	\end{table}
	
	\subsection{Feature Set Construction}
	We are committed to building a feature set based on the statistical characteristics of traffic and adding it to neural network to improve the performance. The URL length of malicious flow, for instance, is generally longer than benign. Experiments indicate that these statistical characteristics are useful for traffic identification.
	
\begin{figure}[htbp]
		\centerline{\includegraphics[scale=1]{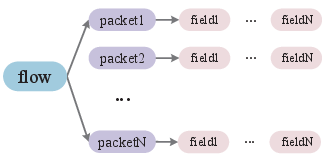}}
		\caption{Hierarchical structure in flow.}
		\label{39}
	\end{figure}

	A flow is represented with corresponding statistical characteristics in dataset BTHT-S. In addition, a flow consists of multiple packets, and a packet consists of different fields. This means that flow can be layered. We show the hierarchical structure in Fig.~\ref{39}. 
	
	\vspace{-1em}
	\vspace{-0.5em}
	
	Traffic in dataset BTHT exhibits this hierarchical timing relationship (such as Fig.~\ref{fig1}). Therefore, we extract features from two statistical levels and form two vectors, packet-level vector ($ PL $) and flow-level vector ($ FL $).
	
	The first is $ PL $. Statistical characteristics of packets are extracted according to items shown in Tab~\ref{tab2x}. RFC1998\cite{chen1996rfc} proposed 47 header fields and suggested that HTTP-based web services use those header fields. Actually, most web services use only a few major header fields. We default that no more than 47 header fields in packets. A packet will generate a extensible vector $ PL(1\times 100) $. The composition and dimension can be adjusted flexibly according to different data Features.
	
	\begin{table}[htbp]
		\caption{Composition of the packet-level vector(PL)}
		\vspace{-1em}
		\begin{center}
			\begin{tabular}{|c|c|}
				\hline
				\textbf{Item Type} & \textbf{Pos} \\
				\hline
				\hline
				request type & 1\\
				\hline
				source port & 2\\
				\hline
				destination port & 3\\
				\hline
				URL length & 4\\
				\hline
				HTTP protocol version & 5\\
				\hline
				field name length & 6-52\\
				\hline
				field value length & 53-99\\
				\hline
				payload & 100\\
				\hline
			\end{tabular}
			\label{tab2x}
		\end{center}
	\end{table}
	
	The second is $ FL $. A flow consists of multiple packets, just as a sentence consists of multiple words. Sometimes, the information provided by a single word is no-valuable and combination of multiple words can show specific content. Similarly, information based on flow level can reflect traffic behavior better. Therefore, we extract statistical characteristics of flows to form vector $FL$. By analyzing the dataset BTHT-R, we default that the number of packets in a flow will not exceed 50. It will be truncated if it is exceeding and filled with 0 if it is insufficient. Tab~\ref{tab3} details items of $FL$. A flow generates a statistical characteristics extensible vector $ FL(1\times 170) $.
	
	\vspace{-1em}
	\begin{table}[htbp]
		\caption{Composition of flow-level vector(FL)}
		\vspace{-1em}
		\begin{center}
			\begin{tabular}{|c|c|}
				\hline
				\textbf{Item Type} & \textbf{Pos} \\
				\hline
				\hline
				number of packets & 1\\
				\hline
				proportion of request packets & 2\\
				\hline
				proportion of response packets & 3\\
				\hline
				proportion of the same packets in the request packets & 4\\
				\hline
				proportion of different packets in the request packets & 5\\
				\hline
				proportion of the same packets in the response packets & 6\\
				\hline
				proportion of different packets in the response packets & 7\\
				\hline
				TTL value of packets & 8-57\\
				\hline
				interval between acquisitions of adjacent messages & 58-106 \\
				\hline
				HTTP packets length & 107\\
				\hline
				proportion of request length & 108\\
				\hline
				proportion of response length & 109\\
				\hline
				length of each packet & 110-159\\
				\hline
				multiple requests corresponding to one response? & 160\\
				\hline
				multiple responses corresponding to one request? & 161\\
				\hline
				'get' proportion in the request & 162\\
				\hline
				'post' proportion in the request & 163\\
				\hline
				'head' proportion in the request & 164\\
				\hline
				other proportions in the request & 165\\
				\hline
				2XX proportion in response & 166\\
				\hline
				4XX proportion in response & 167\\
				\hline
				5XX proportion in response & 168\\
				\hline
				other proportions in response & 169\\
				\hline
				HTTP packets accounted for in flow & 170\\
				\hline
			\end{tabular}
			\label{tab3}
		\end{center}
	\end{table}
	\vspace{-1em}
	
	The $ PL $ and $ FL $ of a flow together constitute a sample in dataset BTHT-S as statistical characteristics of the corresponding raw flow. This feature set can provide more information for the neural network as experience knowledge.
	
	\section{HSTF-Model Methodology}
	
	\subsection{Overview}
	We propose the HSTF-Model (Model based on Hierarchical SpatioTemporal Traffic Features). With a preprocessed HTTP-based flow as input, HSTF-Model can extract features automatically, and judge whether it is malicious or benign.
	
	Combining the layered characteristics of raw flows, HSTF-Model processes data. At packet level, after the feature encoder of raw data outputs the feature matrix, HSTF-Model uses CNN to extract spatial and character features directly. At flow level, HSTF-Model extracts temporal features between packets further by using LSTM. Then, model synthesizes all the feature information through hidden layers, and outputs the discriminant result finally. 
	
	In addition, HSTF-Model is improved by adding statistical feature set. The $ PL $ statistics is added at the packet level and combined with the output of CNN. Then, the $ FL $ statistics is add at the flow level and combine it with the output of LSTM.
	
	The overall structure of HSTF-Model is shown in Fig.~\ref{fig2x}. Through preliminary experiments and empirical knowledge, we determined the overall structure (the number of neurons in each layer of neural network, the output size of feature encoders of $ PL $ and $ FL $, etc.)
	
	The various parts of HSTF-Model are detailed below.
	
	\subsection{Data Preprocessing}
\begin{figure}[htbp]
		\centerline{\includegraphics[scale=0.4]{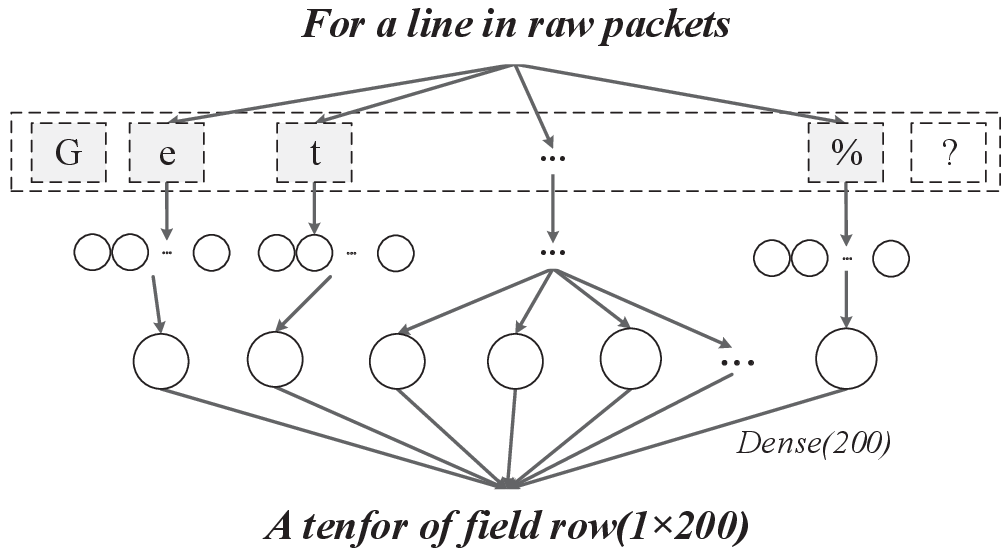}}
		\vspace{-0.5em}
		\caption{Feature encoder of raw data.}
		\label{fig3}
	\end{figure}

	We build feature encoders to preprocess data. The first one is the raw data. As shown in Fig.~\ref{fig3}, the feature encoder of raw data converts each field line in the packet into a $ 1\times 200 $ vector. As a rule of thumb, we default that no more than 47 header fields in packets. Finally, a raw packet produces a $ 47\times 200 $ feature matrix. It will be truncated if it is exceeding and filled with 0 if it is insufficient. Then, we use 1:1 hash technology for data privacy protection and vectorization.
	
	\vspace{-1em}
	\vspace{-0.5em}
	
\begin{figure}[htbp]
		\centerline{\includegraphics[scale=0.5]{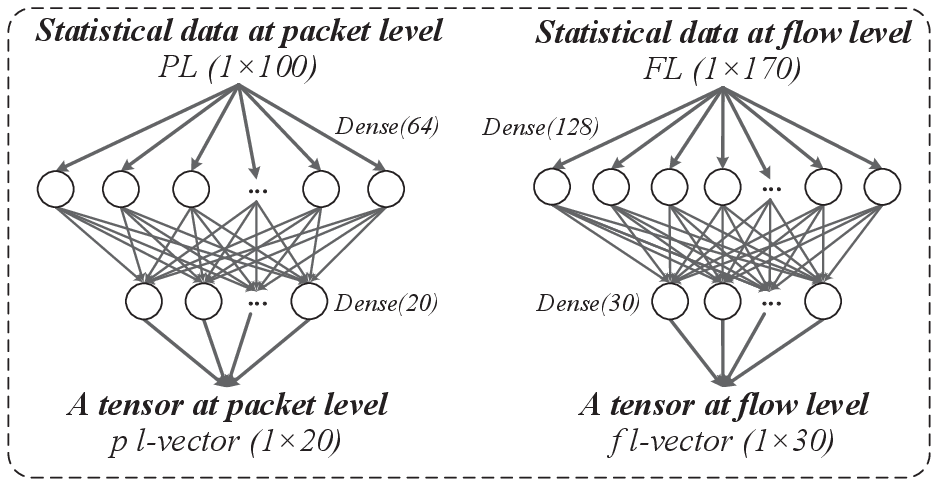}}
		\vspace{-0.5em}
		\caption{Feature encoders of statistical data at packet-level/flow-level.}
		\label{fig4}
	\end{figure}

\begin{figure*}[htbp]
		\centerline{\includegraphics[scale=0.35]{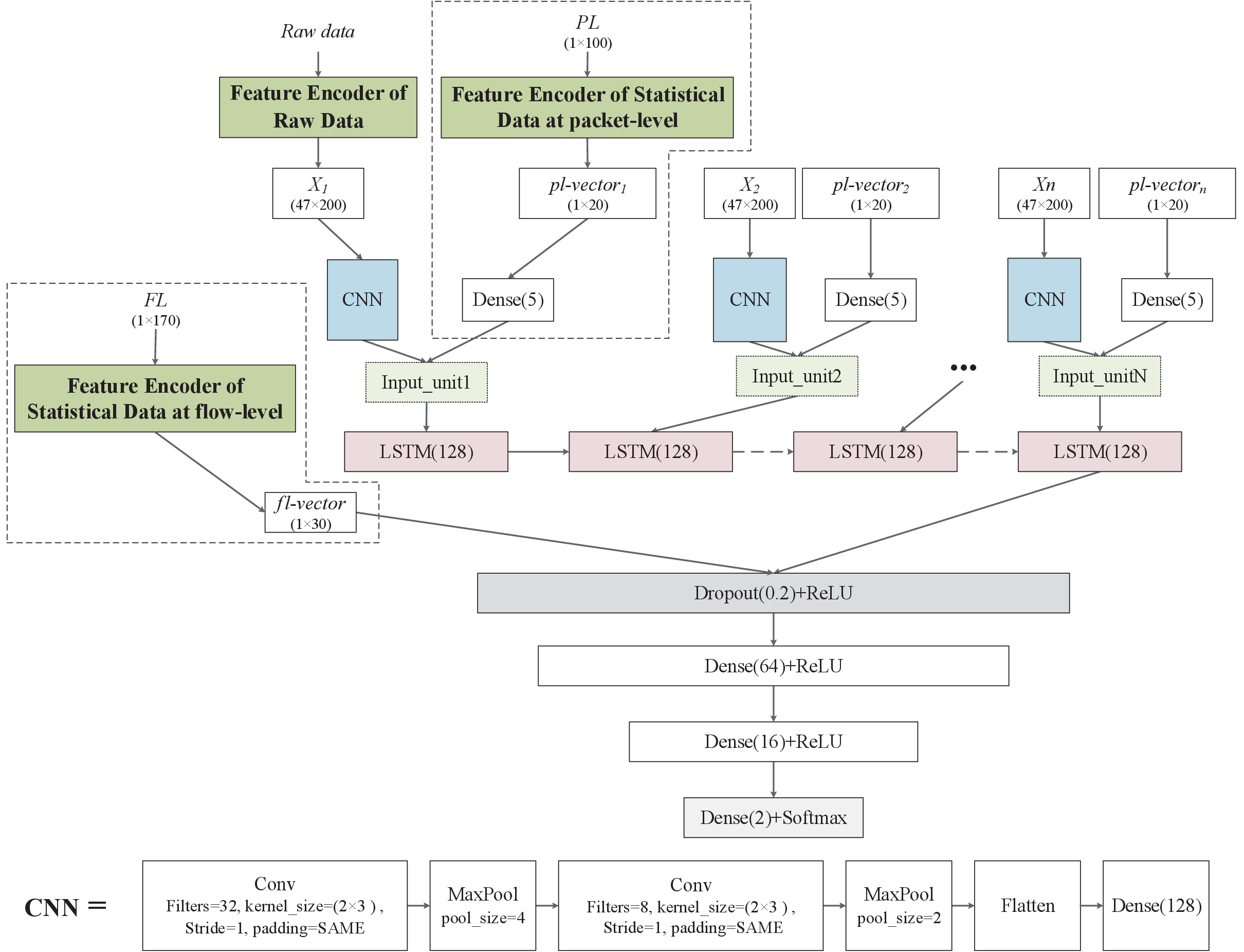}}
		\caption{Overall architecture of HSTF-Model.}
		\label{fig2x}
		\vspace{-1em}
	\end{figure*}

	The second is feature encoders of statistical data at packet-level and flow-level. As shown in Fig.~\ref{fig4}, the feature encoder at packet-level converts $ PL $ to a $ 1\times 20 $ vector and combines it with the output of CNNs. Then, the feature encoder at flow-level converts $ FL $ to a $ 1\times 30 $ vector and combines it with the output of LSTM. Feature encoders convert statistical information into a more compact representation, which facilitates the combination with tensors in neural network.
	
	\vspace{-0.5em}
	\vspace{-1em}
	
	\subsection{Learning of Packet Feature by CNN}
	CNN belongs to deep feedforward neural network. In 1980s, Fukushima \emph{et al.} \cite{Neocognitron} realized it for the first time. CNN can extract hidden structural information. There are one or more convolution and pooling layers in CNN. The convolution layer contains multiple feature mapping neurons (convolution kernels), which separate input data into different feature regions, and each convolution kernel is responsible for extracting local features. We can obtain global feature information of sample by aggregating local features. After convolution operation, the vector is pooled, which is a down-sampling method to reduce the complexity and over-fitting.
	
	We process $ 47\times 200 $ feature matrix by CNN, as shown in Fig.\ref{fig2x}. ReLU is used as an activation function. After two convolutions and pooling operations, the output of CNN is processed by ReLU, and then, enters the follow-up dense layers containing 128 neurons. The output $ C_{i} $ is shown in Eq~\eqref{eq1}, where $ b $ is the bias term and $ f $ is the activation function. Then, multiple convolution kernels perform feature mapping and maximum pooling.
	
	\vspace{-0.5em}
	\begin{equation}
	\begin{aligned}
	c_{i}&=f(w\cdot x_{i:i+h-1}+b)\\
	\hat{c}&=\max(c=\left [ {c_{1} ,c_{2},...,c_{n-h+1}}\right ])
	\label{eq1}
	\end{aligned}
	\end{equation}
	
	\subsection{Learning of Flow Feature by LSTM}
	In 1997, LSTM was proposed by Hochreiter and Schmidhuber\cite{Hochreiter}, which is a variant of RNN. RNN records the processed information previously and makes use of those information in current task. However, there are some disadvantages using RNN. One of them is the difficulty in addressing long-term dependency. It can only utilize the information that is not far from the current task. The LSTM optimizes for this problem by designing 'gate' structures to preserve and select information. Each gate consists of an activation layer and a pointwise operation.
	
	The input gate, \(i_{t}\), combines the input to determine the new information \(\tilde{C_{t}}\) be added. The forgotten gate, \(f_{t}\), handles the previous status and determines the old information \(C_{t-1}\) be discarded. The two gates determine the proportions of new and old information in the current information \(C_{t}\).
	
	\begin{equation}
	\begin{aligned}
	i_{t}&=\delta(W_{i}\cdot \left[h_{t-1}, x_{t}+b_{i}\right] )\\
	\tilde{C_{t}}&=\tanh (W_{C}\cdot \left [h_{t-1}, x_{t} \right]+b_{C})\\
	f_{t}&=\delta (W_{f}\cdot \left [h_{t-1}, x_{t}\right ]+b_{f})\\
	C_{t}&=f_{t}\times C_{t-1}+i_{t}\times \tilde{C_{t}}
	\label{eq2}
	\end{aligned}
	\end{equation}
	
	The output gate, \(o_{t}\), determines the output information of current neurons.
	
	\vspace{-0.9em}
	\begin{equation}
	o_{t}=\delta (W_{o}\cdot \left [h_{t-1}, x_{t}\right ]+b_{o})\label{eq3}
	\end{equation}
	
	There is also a hidden layer status output \(h_{t}\) that is used to assist in the next task processing.
	
	\vspace{-1em}
	\begin{equation}
	h_{t}=o_{t} \times \tanh(C_{t})\label{eq4}
	\end{equation}
	
	CNNs extract different abstract features based on packets in a flow. These features are time-dependent due to the dependence of raw packets. Therefore, we use LSTM with 128 neurons per cell when processing flows in HSTF-Model.
	
	\subsection{Learning of Statistical Characteristics by DNN}
	DNN is one of the foundations of neural network. Rosenblatt proposed the perceptron model in 1958\cite{rosenblatt1958perceptron}, and then derived the multi-layer perceptron (MLP). MLP is also known as simple DNN. In this paper, there is no obvious structural relationship and time dependence in statistical data. Therefore, we process the statistical characteristics by DNN.
	
	A flow including $ n $ packets forms vector of $ n \times 100 + 170 $ size. Then, feature encoders convert $PL$ and $FL$ vector into denser representations through the full connection layers within DNN. At packet level, a dense layer containing 5 neurons further processes the output of feature encoder of $ PL $ for better stitching with the output of CNN. At flow level, the output of feature encoder of $ FL $ is combined with the output of LSTM to enter the follow-up network layer.
	
	\section{Experiment and Evaluation}
	
	\subsection{Evaluation Metrics}
	Precision(P) and recall(R) are calculated to evaluate the performance of HSTF-Model. $ F_{\beta} $ by Eq~\eqref{eq6} is also calculated as the comprehensive evaluation index. We can change the value of $ \beta $ to make the evaluation pay more attention to precision or recall. In this paper, we set $ \beta=1 $, which means that precision and recall are equally important.
	
	\begin{equation}
	\begin{aligned}
	F_{\beta }=\frac{(1+\beta ^{2})\times P\times R}{(\beta ^{2}\times P)+R}
	\end{aligned}\label{eq6}
	\end{equation}
	
	\subsection{Configuration of Environment}
	HSTF-Model is implemented in Python3.5 based on the libraries of Keras and TensorFlow. The system environment of experiments is Ubuntu16.04 LTS. All software applications are deployed on a server machine with 64 CPU cores and 64GB memory. To further accelerate matrix computing, 8 NVIDIA GeForce GTX TITAN X are installed in the server.
	
	For the selection of experimental data, we repeat the experiment 10 times and randomly select 70\% malicious and partial benign flows from dataset BTHT for training at each time. The proportion of malicious:benign is 3:10. In subsequent experiments, the proportion would be changed if necessary. In addition, we select the remaining 30\% malicious flows and 50,000 benign flows from dataset BTHT outside the training set to form the testing set for each experiment.
	
	\begin{table*}[bp]
		\caption{The performance comparison of HSTF-Model with and without statistical characteristics}
		\begin{center}
			\begin{tabular}{|c|c|c|c|c|c|c|c|c|c|c|}
				\hline
				\textbf{Proportion}&\textbf{}&\textbf{}&\multicolumn{4}{|c|}{\textbf{HSTF-Model without statistical characteristics}}&\multicolumn{4}{|c|}{\textbf{HSTF-Model}} \\
				\cline{4-11} 
				\textbf{malicious:benign} & \textbf{Packet size}&\textbf{Flow size}&\textbf{\textit{Precision}}&\textbf{\textit{Recall}}& \textbf{\textit{$ \textbf{F}_{\textbf{1}} $}} &\textbf{Time}&\textbf{\textit{Precision}}& \textbf{\textit{Recall}}& \textbf{\textit{$ \textbf{F}_{\textbf{1}} $}}&\textbf{Time} \\
				\hline
				\hline
				3:10 & 400 & 4 & 99.08 & 98.26 & 98.67 & 11.4s & 99.35 & 99.4 & 99.37 & 11.6s \\
				\hline
				3:10 & 400 & 8 & 99.21 & 99.18 & 99.19 & 13.5 & 99.23 & 99.19 & 99.21 & 15.5s \\
				\hline
				1:4 & 400 & 4 & 99.26 & 96.74 & 97.99 & 11.5s & 99.42 & 97.8 & 98.6 & 11.8s \\
				\hline
				1:4 & 400 & 8 & 99.3 & 97.15 & 98.21 & 13.6s & 99.38 & 98.96 & 99.17 & 16s \\
				\hline
				1:8 & 800 & 8 & 99.65 & 95.43 & 97.49 & 18.4s & 99.94 & 96.64 & 98.26 & 20.8s \\
				\hline
			\end{tabular}
			\label{tabx9}
		\end{center}
	\end{table*}
	
	\subsection{Efficiency of HSTF-Model with Statistical Characteristics}
	In this paper, statistical features containing empirical knowledge are added to the neural network processing as a supplement. These statistics can provide richer data representation and enhance the feature extraction capability of the model. We conduct experiments with different packet size (in bytes) and flow size (in packets) combinations to verify the performance improvement from statistical characteristics. 
	
	We show some experimental results in Tab~\ref{tabx9}. Experiments indicated that the detection effect and robustness of HSTF-Model was generally better than the model without statistical characteristics. But the cost of detection is increased because of the need to extract additional statistical information. However, HSTF-Model is in the same order of magnitude as the complexity of the model without statistical characteristics. Therefore, the cost caused by statistical characteristics is acceptable compared to performance improvements.
	
	In addition, we need to determine the appropriate packet size and flow size, which are not fixed in in different data. The range of packet size and flow size is very wide. We cannot experimentally validate all size combinations. Therefore, the control variable method is used by follow-up experiments.
	
	\subsection{Influence of Packet Size and Flow Size}
\begin{figure}[htbp]
		\centerline{\includegraphics[scale=0.6]{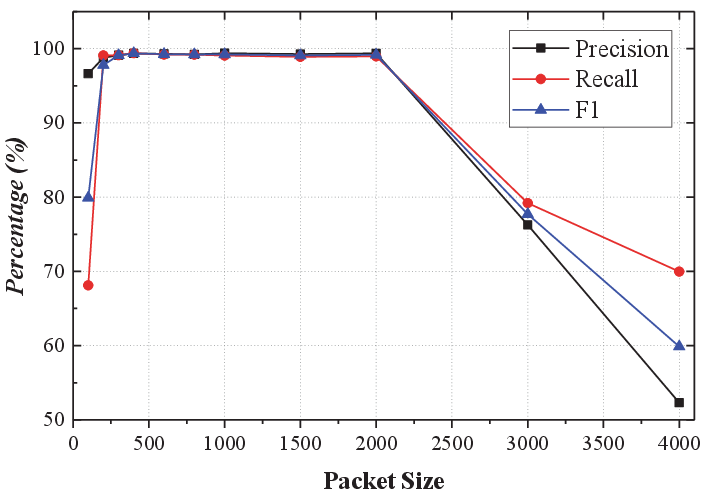}}
		\vspace{-0.5em}
		\caption{Effect of packet size on HSTF-Model (flow size = 4).}
		\label{Doc3}
	\end{figure}

	The appropriate packet size is determined in experiments. According to the statistical analysis of the dataset BTHT and previous experimental experience, we set $flow\ size = 4$. Fig.~\ref{Doc3} details the results.
	
	\vspace{-1em}
	\vspace{-0.5em}
	
	We change the packet size from 100 to 4000 (size = 100, 200, 300, 400, 600, 800, 1000, 1500, 2000, 3000, 4000). HSTF-Model can achieve the best result when $packet\ size=400$ according Fig.~\ref{Doc3}. When the size exceeds 2000, in the experiment, the detection effect begins to decrease. The reason is that most malicious samples in dataset BTHT come from an online package of Trojans, causing practical packet size is not large and the malicious features are basically hidden in preceding bytes. When the size enlarged, it is equivalent to adding noise, which causes the drop-in effect. Therefore, we choose 400 as the packet size for subsequent experiments.
	
\begin{figure}[htbp]
		\centerline{\includegraphics[scale=0.7]{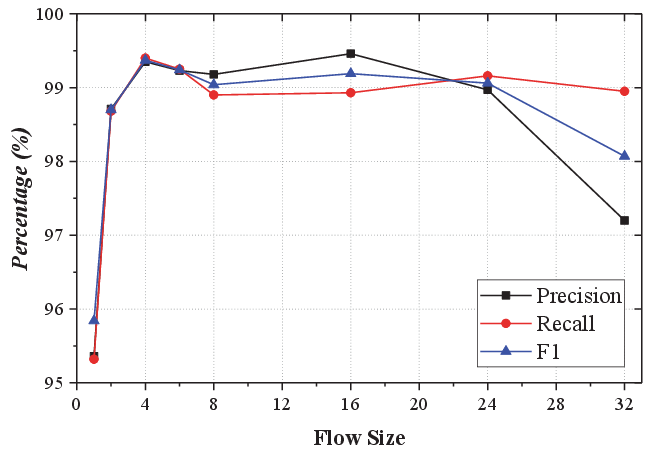}}
		\vspace{-0.5em}
		\caption{Effect of flow size on HSTF-Model (packet size = 400).}
		\label{Doc4}
	\end{figure}

	After CNNs output abstract feature vectors, these vectors form time-series units, which are processed by LSTM. The experimental results of selecting flow size are shown in Fig.~\ref{Doc4}.
	
	\vspace{-0.5em}
	\vspace{-0.5em}
	
	We change the flow size from 1 to 32 (size = 1, 2, 4, 6, 8, 16, 24, 32). HSTF-Model can achieve the best results when $ flow\ size = 4 $. Trojans usually contact the attacker by sending some online packets, which are usually concentrated in the first few packets. Therefore, we only need the first few packets in a flow to determine whether it originated from malicious behavior. In this paper, 4 is chosen as the best flow size.
	
	\subsection{Efficiency of HSTF-Model in Imbalanced Data}
	To evaluate the performance of HSTF-Model in imbalanced data, we use different proportions of data in experiments. The results are shown in Tab~\ref{tab8x}. With the increase in the proportion of benign samples in training, it is easier to extract features of benign data and pays more attention to it. In that case, the training model improves the discriminant threshold for malicious traffic and the precision of the model close to 100\%. For instance, the precision exceeds 99.99\% at proportions of 1:24 and 1:100. However, with the decrease of malicious data, it becomes more difficult to extract the characteristics of malicious samples, and the ability of HSTF-Model to identify malicious samples also declines. At 3:10, the recall reaches 99.4\%, while at 1:100, the recall drops to 78.96\%.
	\vspace{-0.5em}
	\begin{table}[htbp]
		\caption{Efficiency of HSTF-Model at different training proportions}
		\begin{center}
			\begin{tabular}{|c|c|c|c|}
				\hline
				\textbf{Proportion}&\multicolumn{3}{|c|}{\textbf{Evaluation Index$ \textbf{(\%)} $}} \\
				\cline{2-4} 
				\textbf{malicious:benign} & \textbf{\textit{Precision}}& \textbf{\textit{Recall}}& \textbf{\textit{$ \textbf{F}_{\textbf{1}} $}} \\
				\hline
				\hline
				3:10 & 99.35 & 99.4 & 99.37 \\
				\hline
				1:4 & 99.42 & 97.8 & 98.6 \\
				\hline
				1:8 & 99.95 & 96.72 & 98.31 \\
				\hline
				1:24 & 99.99 & 90.83 & 95.19 \\
				\hline
				1:100 & 99.99 & 78.96 & 88.24 \\
				\hline
			\end{tabular}
			\label{tab8x}
		\end{center}
		\vspace{-1.25em}
	\end{table}
	
\begin{figure}[htbp]
		\centerline{\includegraphics[scale=0.8]{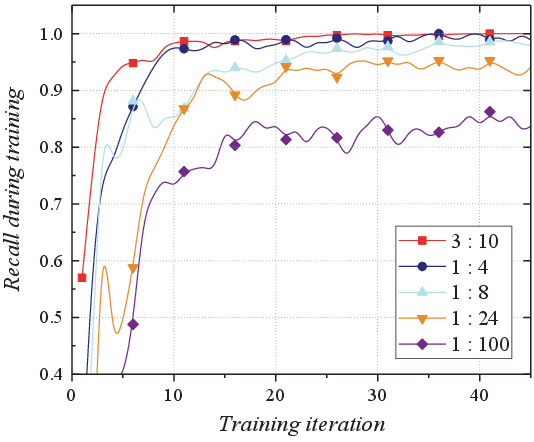}}
		\vspace{-1em}
		\caption{The recall rate convergence curves of HSTF-Model at different training proportions.}
		\label{fig7}
		\vspace{-1.1em}
	\end{figure}

	We show the recall rate convergence process of HSTF-Model at different proportions in Fig.~\ref{fig7}. The performance is getting worse with the decrease of malicious data. But HSTF-Model can still be considered that has good robustness and can handle imbalanced data well. When malicious data only accounts for only 4.17\% in training, it can still converge and has the ${F}_{1}=95.19\%$.
	
	\vspace{-0.5em}

	\subsection{Comparison with Other Methods}
	We implement several of the latest malicious traffic detection methods in combination with our own data analysis. And some classical machine learning algorithms (Bayes, SVM, Decision Tree) are also implemented, which are widely used in the field of malicious traffic detection. We compare HSTF-Model with these methods.
	
	The experimental results are shown in Tab~\ref{tab8}. HSTF-Model has the best comprehensive performance in detection effect and time cost. Rbf-SVM get the recall of only 74.49\% and C4.5 get the precision of only 74.29\%. Although Naive Bayesian costs the least time, its performance is worse than others because it gives up the association between packets. LSTM-R is fast but not excellent because there is no deep consideration of the structural relationship inside the packet. Then, the detection result of S-IDGC and Proposed-Hybrid-Model are excellent but slower than HSTF-Model. These two models are built using traditional machine learning methods, and it is difficult to improve performance with data iterations. HSTF-Model can be continuously updated on the basis of the original. Moreover, HSTF-Model is more robust than other methods when the data is imbalanced in dataset BTHT.
	
	\begin{table}[htbp]
		\caption{HSTF-Model compared to other methods in dataset BTHT}
		\begin{center}
			\begin{tabular}{|c|c|c|c|c|}
				\hline 
				\textbf{Method} & \textbf{\textit{Precision}}& \textbf{\textit{Recall}}& \textbf{\textit{$ \textbf{F}_{\textbf{1}} $}} & \textbf{Test Time}\\
				\hline
				\hline
				Rbf-SVM & 100 & 74.49 & 85.38 & 34m12s \\ 
				\hline
				C4.5 & 74.29 & 98.66 & 84.76 & 1m28s\\ 
				\hline
				GaussianNB & 99.36 & 52.25 & 68.49 & 5.7s\\ 
				\hline
				LSTM-R\cite{radford2018network} & 97.64 & 96.27 & 96.95 & 8.8s\\ 
				\hline
				S-IDGC\cite{chen2018machine} & 99.14 & 99.28 & 99.21 & 3m20s\\ 
				\hline
				\tabincell{c}{Proposed-Hybrid-\\Model\cite{aljawarneh2018anomaly}} & 99.45 & 97.67 & 98.55 & 8m17s\\
				\hline
				\textbf{HSTF-Model} & \textbf{99.35} & \textbf{99.4} & \textbf{99.37} & \textbf{11.6s}\\ 
				\hline
			\end{tabular}
			\label{tab8}
		\end{center}
		\vspace{-1.25em}
	\end{table}
	
	In a word, HSTF-Model has the best comprehensive performance because neural networks have excellent self-learning ability. Meanwhile, statistical characteristics can improve the richness of data, thus enhancing the feature extraction ability of the model and accelerating model convergence.	
	
	\section{Discussion}
	In this paper, we designed the appropriate granularity and range of values to cover the most optimal solutions, and then, determined the optimal values of flow size and packet size by controlling variables. HSTF-Model relies mainly on training data to improve detection capabilities. When the HTTP-based Trojan traffic in the training data is sufficient, the model can still maintain accurate recognition of such Trojan traffic in more complicated traffic environments.
	
	Although HSTF-Model performs well in dataset BTHT, there are still some shortcomings. The first is about false-positives. When a flow is small, there are few features that can be extracted. HSTF-Model is difficult to detect it effectively. The second is that generalization is not fully verified, and the performance of the model is slightly dithered in different situations. For instance, the precision in the real-time Internet will result in a decrease of 2\%-5\% due to the diversity of traffic. Our next work is to alleviate this problem.
	
	\section{Conclusion}
	In this paper, we build features encoders and an effective prototype detection method HSTF-Model. Deep learning is combined with statistical characteristics. It is used for HTTP-based Trojan traffic detection and can reach to 99.4\% in recall (99.35\% in precision) in experiments. In addition, we provide a dataset BTHT consisting of benign traffic and malicious traffic for experiments and other research in the field. 
	
	In the future, we will expand the dataset BTHT further by adding more malicious traffic, and then, enhance the generalization of the model through data iteration. In addition, we only perform a simple coarse-grained partitioning of traffic in this paper, but as the complexity of traffic increases, more fine-grained hierarchical partitioning and feature extraction must be performed . This is one of our next research work.
	
	\section*{Acknowledgment}
	This work was supported by the National Key Research and Development Program of China (No.2016YFB0801502), and the National Natural Science Foundation of China (Grant No.U1736218). The corresponding author of this paper is Shuhao Li.
	\vspace{-0.5em}
	\bibliographystyle{IEEEtran}
	\bibliography{IEEEabrv,Bibliography}
\end{document}